\documentstyle[preprint,aps]{revtex}
\input epsf.tex
\def\DESepsf(#1 width #2){\epsfxsize=#2 \epsfbox{#1}}

\newlength{\dinwidth}
\newlength{\dinmargin}
\begin{document}
\preprint{\vbox{\hbox{hep-ph/0101292}\hbox{YUMS 00-11}}} \draft
\title {Charmless Hadronic Decays of $B$ Mesons \\ to a Pseudoscalar
and a Tensor Meson}
\author{C. S. Kim$^{a,b}$\footnote{kim@kimcs.yonsei.ac.kr,~~~
http://phya.yonsei.ac.kr/\~{}cskim/}, B. H.
Lim$^a$\footnote{bhlim@phya.yonsei.ac.kr} and Sechul
Oh$^a$\footnote{scoh@kimcs.yonsei.ac.kr} }
\address{$^a$ Department of Physics and IPAP, Yonsei University, Seoul,
120-749, Korea \\ $^b$ Department of Physics, University of
Wisconsin, Madison, WI 53706, USA } \maketitle
\begin{abstract}
\noindent We study two-body charmless hadronic decays of $B$
mesons to a pseudoscalar meson ($P$) and a tensor meson ($T$) in
the frameworks of both flavor SU(3) symmetry and generalized
factorization.  Certain ways to test validity of the generalized
factorization are proposed, based on the flavor SU(3) analysis. We
present a set of relations between a flavor SU(3) amplitude and a
corresponding amplitude in the generalized factorization which bridge both
approaches in $B \rightarrow PT$ decays. The branching ratios and
CP asymmetries are calculated using the \emph{full} effective
Hamiltonian including all the \emph{penguin} operators and the
form factors obtained in the non-relativistic quark model of
Isgur, Scora, Grinstein and Wise. We identify the decay modes in
which the branching ratios and CP asymmetries are expected to be
relatively large.
\end {abstract}

\newpage
\section{Introduction}

The CLEO Collaboration has reported new experimental results on
the branching ratios of a number of exclusive $B$ meson decay
modes where $B$ decays into a pair of pseudoscalars ($P$), a
vector ($V$) and a pseudoscalar meson, or a pair of vector mesons.
Motivated by the new data, many works have been done to understand
those exclusive hadronic $B$ decays in the framework of the
generalized factorization, QCD factorization, or flavor SU(3)
symmetry. In the next few years $B$ factories operating at SLAC
and KEK will provide plenty of new experimental data on $B$
decays.  It is expected that improved new bound will be put on the
branching ratios for various decay modes and many decay modes with
small branching ratios will be observed for the first time.  Thus
more information on rare decays of $B$ mesons will be available soon.

There have been a few works \cite{1,2,3} studying two-body
hadronic $B$ decays involving a tensor meson $T$ ($J^P = 2^+$) in
the final state using the non-relativistic quark model of Isgur,
Scora, Grinstein and Wise (ISGW) \cite{4} with the factorization
ansatz. Most of them studied $B$ decays involving a $b \rightarrow
c$ transition, to which only the tree diagram contributes.  In a
recent work \cite{3}, the authors considered the
Cabibbo-Kobayashi-Maskawa (CKM)-suppressed hadronic $B$ decays
involving a $b \rightarrow u$ transition as well as a $b
\rightarrow c$ transition.  However, they included only the tree
diagram contribution even in charmless $B$ decays to $PT$ and
$VT$, such as $B \rightarrow \eta^{(\prime)} a_2$ and $B
\rightarrow \phi f_2^{(\prime)}$.  In most cases of the charmless
$\Delta S =0$ processes, the dominant contribution arises from the
tree diagram and the contributions from the penguin diagrams are
very small. But in some cases such as $B \rightarrow
\eta^{(\prime)} a_2$ and $\eta^{(\prime)} f_2^{(\prime)}$, the
penguin diagrams provide sizable contributions.  Furthermore, in
the charmless $|\Delta S| =1$ decay processes, the penguin diagram
contribution is enhanced by the CKM matrix elements $V^*_{tb}
V_{ts}$ and becomes dominant.

Experimentally several tensor mesons have been observed \cite{4A},
such as the isovector $a_2$(1320), the isoscalars $f_2$(1270),
$f_2^{\prime}$(1525), $f_2$(2010), $f_2$(2300), $f_2$(2340),
$\chi_{c2}(1P)$, $\chi_{b2}(1P)$ and $\chi_{c2}(2P)$, the
isospinor $K_2^*$(1430) and $D_2^*$(2460).  Experimental data on
the branching ratios for $B$ decays involving a tensor meson in
the final state provide  only upper bounds, as follows \cite{4A}:
\begin{eqnarray}
{\cal B} (B^{+(0)} \rightarrow D_2^*(2460)^{0(-)} \pi^+) &<& 1.3
(2.2) \times 10^{-3},   \nonumber
\\ {\cal B} (B^{+(0)} \rightarrow D_2^*(2460)^{0(-)} \rho^+) &<& 4.7
(4.9) \times 10^{-3},   \nonumber
\\ {\cal B} (B^+ \rightarrow K_2^*(1430)^0 \pi^+) &<& 6.8 \times
10^{-4},   \nonumber
\\ {\cal B} (B^{+(0)} \rightarrow K_2^*(1430)^{+(0)} \rho^0) &<& 1.5
(1.1) \times 10^{-3},   \nonumber
\\ {\cal B} (B^{+(0)} \rightarrow K_2^*(1430)^{+(0)} \phi) &<& 3.4
(1.4) \times 10^{-3},   \nonumber
\\ {\cal B} (B^+ \rightarrow \pi^+ f_2(1270)) &<& 2.4 \times
10^{-4},   \nonumber
\\ {\cal B} (B^+ \rightarrow \rho^0 a_2(1320)^+) &<& 7.2 \times
10^{-4},   \nonumber
\\ {\cal B} (B^0 \rightarrow \pi^{\pm} a_2(1320)^{\mp}) &<& 3.0 \times
10^{-4}.
\end{eqnarray}

In this work, we analyze two-body charmless hadronic decays of $B$
mesons to a pseudoscalar meson and a tensor meson in the
frameworks of \emph{both} flavor SU(3) symmetry and generalized
factorization.  Purely based on the flavor SU(3) symmetry, we
first present a model-independent analysis in $B \rightarrow PT$
decays.  Then we use the \emph{full} effective Hamiltonian
including all the penguin operators and the ISGW quark model to
calculate the branching ratios for $B \rightarrow PT$ decays.
Since we include both the tree and the penguin diagram
contributions to decay processes, we are able to calculate the
branching ratios for all the charmless $|\Delta S| =1$ decays and
the relevant CP asymmetries.  In order to bridge the flavor SU(3)
approach and the factorization approach, we present a set of
relations between a flavor SU(3) amplitude and a corresponding
amplitude in factorization in $B \rightarrow PT$ decays. Certain
ways to test validity of the generalized factorization are
proposed by emphasizing interplay between both approaches.

We organize this work as follows.  In Sec. II we discuss the
notations for SU(3) decomposition and the full effective
Hamiltonian for $B$ decays.  In Sec. III we present a model-
independent analysis of $B \rightarrow PT$ decays based on SU(3)
symmetry.  In Sec. IV the two-body decays $B \rightarrow PT$ are
analyzed in the framework of generalized factorization.  The
branching ratios and CP asymmetries are calculated using the form
factors obtained in the ISGW quark model.  Finally, in Sec. V our
results are summarized.

\section{Framework}

In the flavor SU(3) approach, the decay amplitudes of two-body $B$
decays are decomposed into linear combinations of the SU(3)
amplitudes, which are reduced matrix elements defined in Ref.
\cite{5}.  In SU(3) decomposition of decay amplitudes of the $B
\rightarrow PT$ processes, we choose the notations given in Refs.
\cite{5,7,8} as follows: We represent the decay amplitudes in
terms of the basis of quark diagram contributions, $T$ (tree), $C$
(color-suppressed tree), $P$ (QCD-penguin), $S$ (additional
penguin effect involving SU(3)-singlet mesons), $E$ (exchange),
$A$ (annihilation), and {\it PA} (penguin annihilation). The
amplitudes $E$, $A$ and {\it PA} may be neglected to a good
approximation because of a suppression factor of $f_B / m_B
\approx 5 \%$.  For later convenience we also denote the
electroweak (EW) penguin effects explicitly as $P_{EW}$
(color-favored EW penguin) and $P_{EW}^{C}$ (color-suppressed EW
penguin), even though in terms of quark diagrams the inclusion of
these EW penguin effects only leads to the following replacement
without introducing new SU(3) amplitudes; $T \rightarrow T +
P_{EW}^{C}$, $C \rightarrow C + P_{EW}$, $P \rightarrow P -{1
\over 3} P_{EW}^{C}$, $S \rightarrow S -{1 \over 3} P_{EW}$.  The
phase convention used for the pseudoscalar and the tensor mesons
is
\begin{eqnarray}
\pi^+ (a^+_2) &=& u \bar d~, \;\;\; \pi^0 (a^0_2) = -{1 \over
\sqrt{2}} (u \bar u -d \bar d)~, \;\;\; \pi^- (a^-_2) = -\bar u d~,
\nonumber
\\ K^+ (K^{*+}_2) &=& u \bar s~, \;\;\; K^0 (K^{*0}_2) = d \bar s~,
\;\;\; \bar K^0 (\bar K^{*0}_2) = \bar d s~, \;\;\; K^- (K^{*-}_2)
= - \bar u s~,  \nonumber
\\ \eta &=& -{1 \over \sqrt{3}} (u \bar u + d \bar d - s \bar s)~,
\;\; \eta^{\prime} = {1 \over \sqrt{6}} (u \bar u + d \bar d +2 s
\bar s)~,   \nonumber
\\ f_2 &=& {1 \over \sqrt{2}} (u \bar u +d \bar d) \cos \phi_{_T} +(s
\bar s) \sin \phi_{_T} ~, \;\; f^{\prime}_2 = {1 \over \sqrt{2}}
(u \bar u +d \bar d) \sin \phi_{_T} -(s \bar s) \cos \phi_{_T} ~,
\end{eqnarray}
where the mixing angle $\phi_{_T}$ is given by $\phi_{_T} =
\arctan(1 / \sqrt{2}) - 28^0 \approx 7^0$ \cite{1,8A}.

In the factorization scheme, we first consider the effective weak
Hamiltonian. We then use the generalized factorization
approximation to derive hadronic matrix elements by saturating the
vacuum state in all possible ways.  The method includes color
octet non-factorizable contribution by treating $\xi\equiv 1/N_c$
($N_c$ denotes the effective number of color) as an adjustable
parameter.  The generalized factorization approximation has been
quite successfully used in two-body $D$ decays as well as $B
\rightarrow D$ decays\cite{9}. The effective weak Hamiltonian for
hadronic $\Delta B=1$ decays can be written as
\begin{eqnarray}
 H_{eff} &=& {4 G_{F} \over \sqrt{2}} \left[ V_{ub}V^{*}_{uq} (c_1
O^{u}_1 +c_2 O^{u}_2)
   + V_{cb}V^{*}_{cq} (c_1 O^{c}_1 +c_2 O^{c}_2)
   - V_{tb}V^{*}_{tq} \sum_{i=3}^{12} c_{i} O_{i} \right] \nonumber \\
  &+& {\rm H.C.} ~,
\end{eqnarray}
where $O_{i}$'s are defined as
\begin{eqnarray}
O^f_1 &=& (\bar q \gamma_{\mu} L f) (\bar f \gamma^{\mu} L b) ~,
\;\; O^f_2 = (\bar q_{\alpha} \gamma_{\mu} L f_{\beta}) (\bar
f_{\beta} \gamma^{\mu} L b_{\alpha})~,    \nonumber \\ O_{3(5)} &=&
(\bar q \gamma_{\mu} L b) (\Sigma \bar q^{\prime} \gamma^{\mu}
L(R) q^{\prime})~,  \;\; O_{4(6)} = (\bar q_{\alpha} \gamma_{\mu} L
b_{\beta}) (\Sigma \bar q^{\prime}_{\beta} \gamma^{\mu} L(R)
q^{\prime}_{\alpha})~,  \nonumber \\ O_{7(9)} &=& {3 \over 2} (\bar
q \gamma_{\mu} L b) (\Sigma e_{q^{\prime}} \bar q^{\prime}
\gamma^{\mu} R(L) q^{\prime})~,  \;\; O_{8(10)} ={3 \over 2} (\bar
q_{\alpha} \gamma_{\mu} L b_{\beta}) (\Sigma e_{q^{\prime}} \bar
q^{\prime}_{\beta} \gamma^{\mu} R(L) q^{\prime}_{\alpha})~,
\nonumber \\ O_{11} &=& {g_s \over 32 \pi^2} m_b (\bar q
\sigma^{\mu \nu} R T^a b) G^a_{\mu \nu}~,  \;\; O_{12} = {e \over
32 \pi^2} m_b (\bar q \sigma^{\mu \nu} R b) F_{\mu \nu}~.
\end{eqnarray}
Here $c_i$'s are the Wilson coefficients (WC's) evaluated at the
renormalization scale $\mu$.  And $L(R) = (1 \mp \gamma_5)/2$, $f$
can be $u$ or $c$ quark, $q$ can be $d$ or $s$ quark, and
$q^{\prime}$ is summed over $u$, $d$, $s$, and $c$ quarks.
$\alpha$ and $\beta$ are the SU(3) color indices, and $T^a$ ($a=
1,...,8$) are the SU(3) generator with the normalization $Tr (T^a
T^b) =\delta^{ab} /2$.  $g_s$ and $e$ are the strong and electric
couplings, respectively.  $G^a_{\mu \nu}$ and $F_{\mu \nu}$ denote
the gluonic and photonic field strength tensors, respectively.
$O_1$ and $O_2$ are the tree-level and QCD-corrected operators.
$O_{3-6}$ are the gluon-induced strong penguin operators.
$O_{7-10}$ are the EW penguin  operators due to $\gamma$ and $Z$
exchange, and box diagrams at loop level.  We shall take into
account the chromomagnetic operator $O_{11}$ but neglect the
extremely small contribution from $O_{12}$. The dipole
contribution is in general quite small, and is of the order of
$10\%$ for penguin dominated modes. For all the other modes it can
be neglected \cite{10}.

We use the ISGW quark model to analyze two-body charmless decay
processes $B \rightarrow PT$  in the framework of generalized
factorization.  We describe the parameterizations of the hadronic
matrix elements in $B \rightarrow PT$ decays: \cite{4}
\begin{eqnarray}
\langle 0 | A^{\mu} | P \rangle &=& i f_P p_P^{\mu} ~,  \\
\langle T | j^{\mu} | B \rangle &=& i h(m_P^2) \epsilon^{\mu \nu
\rho \sigma} \epsilon^*_{\nu \alpha} p_B^{\alpha} (p_B
+p_T)_{\rho} (p_B -p_T)_{\sigma} + k(m_P^2) \epsilon^{* \mu \nu}
(p_B)_{\nu}  \nonumber \\
&\mbox{}&  + \epsilon^*_{\alpha \beta} p_B^{\alpha} p_B^{\beta} [
b_+(m_P^2) (p_B +p_T)^{\mu} +b_-(m_P^2) (p_B -p_T)^{\mu} ]~,
\label{formfactor}
\end{eqnarray}
where $j^{\mu} = V^{\mu} -A^{\mu}$.  $V^{\mu}$ and $A^{\mu}$
denote a vector and an axial-vector current, respectively.  $f_P$
denotes the decay constant of the relevant pseudoscalar meson.
$h(m_P^2)$, $k(m_P^2)$, $b_+(m_P^2)$, and $b_-(m_P^2)$ express the
form factors for the $B \rightarrow T$ transition,
$F^{B \rightarrow T}(m_P^2)$, which have been
calculated at $q^2 =m_P^2$ $(q^{\mu} \equiv p_B^{\mu} -p_T^{\mu})$
in the ISGW quark model \cite{4}. $p_B$ and $p_T$ denote the
momentum of the $B$ meson and the tensor meson, respectively. We
note that the matrix element
\begin{equation}
\langle 0 | j^{\mu} | T \rangle =0~,
\label{fT}
\end{equation}
because the trace of the polarization tensor
$\epsilon^{\mu \nu}$ of the tensor meson $T$ vanishes and the
auxiliary condition holds, $p_T^{\mu} \epsilon_{\mu \nu} =0$
\cite{1}.  Thus, in the generalized factorization scheme,
the decay amplitudes
for $B \rightarrow PT$ can be considerably simplified, compared to those
for other two-body charmless decays of $B$ mesons such as $B
\rightarrow PP$, $PV$, and $VV$: Any decay amplitude for $B
\rightarrow PT$ is simply proportional to the decay constant $f_P$
and a certain linear combination of the form factors $F^{B \rightarrow T}$,
{\it i.e.}, there is no such amplitude proportional to
$f_T \times F^{B \rightarrow P}$.

\section{Flavor SU(3) analysis of $B \rightarrow PT$ decays}

We list the $B \rightarrow PT$ decay modes in terms of the SU(3)
amplitudes.  The coefficients of the SU(3) amplitudes in $B
\rightarrow PT$ are listed in Tables I and II for
strangeness-conserving ($\Delta S =0$) and strangeness-changing
($|\Delta S| =1$) processes, respectively. In the tables, the
unprimed and the primed letters denote $\Delta S =0$ and
$|\Delta S| =1$ processes, respectively.
The subscript, $P$ in $T_P,C_P,$ ... or $T$ in $T_T,C_T,$ ..., on each
SU(3) amplitude is used to describe such a case that the meson, which
includes the spectator quark in the corresponding quark diagram, is the
pseudoscalar $P$ or the tensor $T$.  Note that the coefficients of the
SU(3) amplitudes with the subscript $P$, which would be proportional to
$f_T \times F^{B \rightarrow P}$, are
expressed in square brackets.  As explained in Sec. II, the contributions of
the SU(3) amplitudes with the subscript $P$ vanish in the framework of
factorization, because those contributions contain the matrix
element $\langle T| J^{\rm{weak}}_{\mu} |0 \rangle$ which is zero,
see Eq. (\ref{fT}).
Thus, it will be interesting to compare
the results obtained in the SU(3) analysis with those obtained in
the factorization scheme, as we shall see.  We will present some
ways to test validity of both schemes in future experiment.

Among the $\Delta S =0$ amplitudes, the tree diagram contribution
is expected to be largest so that from Table I the decays $B^+
\rightarrow \pi^+ a_2^0$, $\pi^+ f_2$, and $B^0 \rightarrow \pi^+
a_2^-$ are expected to have the largest rates.  Here we have
noticed that in $B^+ \rightarrow \pi^+ f_2^{(\prime)}$ decays,
$\cos \phi_{_T} =0.99$ and $\sin \phi_{_T} =0.13$, since the
mixing angle $\phi_{_T} \approx 7^0$. The amplitudes for the
processes $B \rightarrow K K^*_2$ have only penguin diagram
contributions, and so they are expected to be small.  In
principle, the penguin contribution (combined with the smaller
color-suppressed EW penguin) $p_T \equiv P_T^{\prime} -{1 \over 3}
P_{EW,T}$ can be measured in $B^{+(0)} \rightarrow \bar K^0
K_2^{*+(0)}$.  The tree contribution (combined with much smaller
color-suppressed EW penguin) $t_T \equiv T_T +P^C_{EW,T}$ are
measured by the combination $A(B^{+(0)} \rightarrow \bar K^0
K_2^{*+(0)}) -A(B^0 \rightarrow \pi^+ a_2^-)$.  The amplitudes for
$B^0 \rightarrow \pi^0 f_2^{\prime}$, $\eta f_2^{\prime}$, and
$\eta^{\prime} f_2^{\prime}$ have the color-suppressed tree
contributions, $C_T (C_P)$, but are suppressed by $\sin \phi$ so
that they are expected to be small. We shall see that these
expectations based on the SU(3) approach are consistent with those
calculated in the factorization approximation. However, there
exist some cases in which the predictions based on both approaches
are inconsistent. Note that in Table I the amplitudes for $B^0
\rightarrow \pi^- a_2^+$ and $B^{+ (0)} \rightarrow K^{+(0)} \bar
K^{*0}_2$ can be decomposed into linear combinations of the SU(3)
amplitudes as follows:
\begin{eqnarray}
A(B^0 \rightarrow \pi^- a_2^+) &=& - T_P -P_P -(2/3) P^C_{EW,P}~,
\label{b0pia2}  \\
A(B^+ \rightarrow K^+ \bar K^{*0}_2) &=& A(B^0 \rightarrow K^0
\bar K^{*0}_2) = P_P -(1/3) P^C_{EW,P}~.
\end{eqnarray}
As previously explained, in factorization the rates for these
processes vanish because all the SU(3) amplitudes are with the subscript $P$.
Non-zero of decay rates for these processes would arise from non-factorizable
effects or final state interactions. Thus, in principle one can test validity
of the factorization ansatz by measuring the rates for these decays in
future experiment.  Therefore, the non-factorizable penguin contribution,
if exists, (combined with the smaller color-suppressed EW penguin) $p_P
\equiv P_P -{1 \over 3} P_{EW,P}$ can be measured in $B^{+(0)}
\rightarrow \bar K^{+(0)} \bar K_2^{*+(0)}$.  Also, supposing that
$P_P$ is very small compared to $T_P$ as usual, one can determine
the magnitude of $T_P$ by measuring the rate for $B^0 \rightarrow
\pi^- a^+_2$.

In the $|\Delta S|=1$ decays, the (strong) penguin contribution
$P^{\prime}$ is expected to dominate because of enhancement by the
ratio of the CKM elements $|V^*_{tb} V_{ts}| / |V^*_{ub} V_{us}|
\approx 50$.  We note that the amplitudes for $B^+ \rightarrow K^0
a_2^+$ and $B^+ \rightarrow \pi^+ K_2^{*0}$ have only penguin
contributions, respectively, as follows:
\begin{eqnarray}
A (B^+ \rightarrow K^0 a_2^+) &=& P_T^{\prime} -{1 \over 3}
P_{EW,T}^{C \prime}~, \\
A (B^+ \rightarrow \pi^+ K_2^{*0}) &=& P_P^{\prime} -{1 \over 3}
P_{EW,P}^{C \prime}~. \label{piK2}
\end{eqnarray}
Thus the penguin contribution (combined with the smaller
color-suppressed EW penguin) $p_T^{\prime} \equiv P_T^{\prime} -{1
\over 3} P_{EW,T}^{C \prime}$ is measured in $B^+ \rightarrow K^0
a_2^+$.  Similarly, $p_P^{\prime} \equiv P_P^{\prime} -{1 \over 3}
P_{EW,P}^{C \prime}$ is determined in $B^+ \rightarrow \pi^+
K_2^{*0}$. (In fact, $p_P^{\prime} =0$ in factorization.)  By
comparing the branching ratios for these two modes measured in
experiment, one can determine which contribution (i.e.,
$p_T^{\prime}$ or $p_P^{\prime}$) is larger. The (additional
penguin) SU(3) singlet amplitude $S^{\prime}$ is expected to be
very small because of the Okubo-Zweig-Iizuka (OZI) suppression,
but the SU(3) singlet amplitude $S^{\prime}$ for the decays
involving the pseudoscalar mesons $\eta$ and $\eta^{\prime}$ is
expected not to be very small, since the flavor-singlet couplings
of the $\eta$ and $\eta^{\prime}$ can be affected by the axial
anomaly\cite{11}. Thus, from Table II, one can expect that the
processes $B^{+(0)} \rightarrow \eta^{\prime} K^{*+(0)}_2$ have
large branching ratios compared to other $|\Delta S|=1$ decays,
since they have both of the penguin contributions $P^{\prime}$ and
$S^{\prime}$ (and the smaller EW penguin contributions
$P^{\prime}_{EW}$ and $P^{C \prime}_{EW}$) and these contributions
interfere constructively such as $2 P^{\prime}_T +P^{\prime}_P +4
S^{\prime}_T$. In contrast, the processes $B^{+(0)} \rightarrow
\eta K^{*+(0)}_2$ have both of the penguin contributions
$P^{\prime}$ and $S^{\prime}$, but these interfere destructively
such as $- P^{\prime}_T +P^{\prime}_P +S^{\prime}_T$.  As in
$\Delta S=0$ decays, there are certain processes whose amplitudes
can be expressed by the SU(3) amplitudes, but are expected to
vanish in factorization:  For instance, $A (B^+ \rightarrow \pi^+
K_2^{*0})$ is given by Eq. (\ref{piK2}) and $A (B^0 \rightarrow
\pi^- K_2^{*+}) = -(T_P^{\prime} +P_P^{\prime} +{2 \over 3}
P_{EW,P}^{C \prime})$.  Thus, in principle measurement of the
rates for these decays can be used to test the factorization
ansatz.  We also note that the decay amplitudes for modes $B^+
\rightarrow \pi^0 K_2^{*+}$ and $B^0 \rightarrow \pi^0 K_2^{*0}$
can be respectively written as
\begin{eqnarray}
A (B^+ \rightarrow \pi^0 K_2^{*+}) &=& -{1 \over \sqrt{2}}
(T_P^{\prime} + C_T^{\prime} +P_P^{\prime} +P_{EW,T}^{\prime} +{2
\over 3} P_{EW,P}^{C \prime})~, \\
A (B^0 \rightarrow \pi^0 K_2^{*0}) &=& -{1 \over \sqrt{2}} (
C_T^{\prime} -P_P^{\prime} +P_{EW,T}^{\prime} +{1 \over 3}
P_{EW,P}^{C \prime})~.
\end{eqnarray}
Since in factorization only the amplitudes having the subscript
$T$ does not vanish, we shall see that ${\cal B} (B^+ \rightarrow
\pi^0 K_2^{*+}) = {\cal B} (B^0 \rightarrow \pi^0 K_2^{*0})$ in
the factorization scheme, where ${\cal B}$ denotes the branching
ratio.  Thus, if $P_P^{\prime}$ or $T_P^{\prime}$ is (not zero
and) not very suppressed compared to $C_T^{\prime}$, then there
would be a sizable discrepancy in the relation
${\cal B} (B^+ \rightarrow \pi^0 K_2^{*+})
= {\cal B} (B^0 \rightarrow \pi^0 K_2^{*0})$, and in
principle it can be tested in experiment.

{}From Tables I and II, we find some useful relations among the
decay amplitudes.  The equivalence relations are
\begin{eqnarray}
A(B^+ \rightarrow K^+ \bar K^{*0}_2) &=& A(B^0 \rightarrow K^0
\bar K^{*0}_2)~, \nonumber
\\ A(B^+ \rightarrow \bar K^0 K^{*+}_2) &=& A(B^0 \rightarrow \bar
K^0 K^{*0}_2)~.
\end{eqnarray}
The quadrangle relations are: for the $\Delta S = 0$ processes,
\begin{eqnarray}
\sqrt{2} A(B^+ \rightarrow \eta^{\prime} a^+_2) + A(B^+
\rightarrow \eta a^+_2) &=& 2 A(B^0 \rightarrow \eta^{\prime}
a^0_2) +\sqrt{2} A(B^0 \rightarrow \eta a^0_2)~,  \nonumber
\\
{1 \over c} [ A(B^+ \rightarrow \pi^+ f_2) -\sqrt{2} A(B^0
\rightarrow \pi^0 f_2)] &=& {1 \over s} [ A(B^+ \rightarrow \pi^+
f^{\prime}_2) -\sqrt{2} A(B^0 \rightarrow \pi^0 f^{\prime}_2)]
\nonumber
\\ = {1 \over c} [ \sqrt{2} A(B^0 \rightarrow \eta^{\prime} f_2) +
A(B^0 \rightarrow \eta f_2)] &=& {1 \over s} [ \sqrt{2} A(B^0
\rightarrow \eta^{\prime} f^{\prime}_2) + A(B^0 \rightarrow \eta
f^{\prime}_2)]~,    \label{deltas0}
\end{eqnarray}
and for the $|\Delta S| = 1$ processes,
\begin{eqnarray}
\sqrt{2} A(B^+ \rightarrow K^+ a^0_2) + A(B^+ \rightarrow K^0
a^+_2) &=& A(B^0 \rightarrow K^+ a^-_2) + \sqrt{2} A(B^0
\rightarrow K^0 a^0_2)~,  \nonumber
\\ {1 \over c} [ A(B^+ \rightarrow K^+ f_2)
- A(B^0 \rightarrow K^0 f_2)] &=& {1 \over s} [ A(B^+ \rightarrow
K^+ f^{\prime}_2) - A(B^0 \rightarrow K^0 f^{\prime}_2)]~,
\nonumber
\\ A(B^+ \rightarrow \pi^+ K^{*0}_2) +\sqrt{2} A(B^+ \rightarrow \pi^0
K^{*+}_2) &=& A(B^0 \rightarrow \pi^- K^{*+}_2) +\sqrt{2} A(B^0
\rightarrow \pi^0 K^{*0}_2)~,  \nonumber
\\ A(B^+ \rightarrow \eta K^{*+}_2)
+\sqrt{2} A(B^+ \rightarrow \eta^{\prime} K^{*+}_2) &=& A(B^0
\rightarrow \eta K^{*0}_2) +\sqrt{2} A(B^0 \rightarrow
\eta^{\prime} K^{*0}_2)~,  \label{deltas1}
\end{eqnarray}
where $c \equiv \cos \phi_{_T}$ and $s \equiv \sin \phi_{_T}$.
Note that the above relations are derived, purely based on flavor
SU(3) symmetry.  In the factorization scheme, (neglecting the
SU(3) amplitudes with the subscript $P$) we would have in addition
the approximate relations as follows.\footnote{Considering SU(3)
breaking effects, we use the symbol $\approx$ in the following
relations instead of the equivalence symbol $=$.} The following
factorization relation would hold:
\begin{eqnarray}
\sqrt{2} A(B^+ \rightarrow \pi^+ a_2^0) &\approx& A(B^0
\rightarrow \pi^+ a_2^-)~.  \label{facto1}
\end{eqnarray}
The quadrangle relations given in Eqs. (\ref{deltas0},
\ref{deltas1}) would be divided into the following factorization
relations; for the $\Delta S = 0$ processes,
\begin{eqnarray}
A(B^+ \rightarrow \eta a^+_2) &\approx& \sqrt{2} A(B^0 \rightarrow
\eta a^0_2)   \nonumber
\\ \approx -\sqrt{2} A(B^+ \rightarrow \eta^{\prime} a^+_2) &\approx&
-2 A(B^0 \rightarrow \eta^{\prime} a^0_2)~,  \nonumber
\\ {1 \over c} A(B^{+(0)} \rightarrow \pi^{+(0)} f_2) &\approx& {1
\over s} A(B^{+(0)} \rightarrow \pi^{+(0)} f^{\prime}_2)~,
\nonumber
\\ {1 \over c} A(B^0 \rightarrow \eta f_2) &\approx& {1 \over s}
A(B^0 \rightarrow \eta f^{\prime}_2)   \nonumber
\\ \approx -{1 \over c} \sqrt{2} A(B^0 \rightarrow \eta^{\prime} f_2)
&\approx& -{1 \over s} \sqrt{2} A(B^0 \rightarrow \eta^{\prime}
f^{\prime}_2)~,  \label{facto2}
\end{eqnarray}
and for the $|\Delta S| = 1$ processes,
\begin{eqnarray}
\sqrt{2} A(B^+ \rightarrow K^+ a_2^0) &\approx& A(B^0 \rightarrow
K^+ a_2^-)~,  \nonumber
\\ A(B^+ \rightarrow K^0 a^+_2) &\approx& \sqrt{2} A(B^0 \rightarrow
K^0 a^0_2)~,  \nonumber
\\ {1 \over c} A(B^+ \rightarrow K^+ f_2) &\approx& {1 \over s} A(B^+
\rightarrow K^+ f^{\prime}_2)~,  \nonumber
\\ {1 \over c} A(B^0 \rightarrow K^0 f_2) &\approx& {1 \over s}
A(B^0 \rightarrow K^0 f^{\prime}_2)~,  \nonumber
\\ A(B^+ \rightarrow \pi^0 K^{*+}_2) &\approx&
A(B^0 \rightarrow \pi^0 K^{*0}_2)~,  \nonumber
\\ A(B^+ \rightarrow \eta K^{*+}_2) &\approx& A(B^0
\rightarrow \eta K^{*0}_2)~,  \nonumber
\\ A(B^+ \rightarrow \eta^{\prime} K^{*+}_2) &\approx& A(B^0
\rightarrow \eta^{\prime} K^{*0}_2)~.  \label{facto3}
\end{eqnarray}
Therefore, in principle the above relations given in Eqs.
(\ref{facto1}, \ref{facto2}, \ref{facto3}) provide an interesting
way to test the factorization scheme by measuring and comparing
magnitudes of the decay amplitudes involved in the relations.  In
consideration of SU(3) breaking effects, the relation in Eq.
(\ref{facto1}) is best to use, because in fact the relation arises
from isospin symmetry assuming $C_P = P_P = P_{EW,P} =P^C_{EW,P}
=0$.  (However, if $C_P$ is negligibly small (though not zero)
compared to $T_T$, Eq. (\ref{facto1}) will approximately hold.)

\section{Analysis of $B \rightarrow PT$ decays using the Isgur-Scora-
Grinstein-Wise model}

Now, we present expressions for SU(3) amplitudes involved in $B
\rightarrow PT$ decays as calculated in the factorization scheme
as follows \cite{12}.  (Note that all the SU(3) amplitudes with
the subscript $P$ vanish because those are proportional to the
matrix element $\langle T | j^{\mu} | 0 \rangle$.)
\begin{eqnarray}
T^{(\prime)}_T &=& i {G_F \over \sqrt{2}} V^*_{ub} V_{ud (s)} (f_P
\epsilon^*_{\mu \nu} p^{\mu}_B p^{\nu}_B F^{B \rightarrow
T}(m_P^2)) a_1~, \nonumber \\
C^{(\prime)}_T &=& i {G_F \over \sqrt{2}} V^*_{ub} V_{ud (s)} (f_P
\epsilon^*_{\mu \nu} p^{\mu}_B p^{\nu}_B F^{B \rightarrow
T}(m_P^2)) a_2~,  \nonumber \\
S^{(\prime)}_T &=& -i {G_F \over \sqrt{2}} V^*_{tb} V_{td (s)}
(f_P \epsilon^*_{\mu \nu} p^{\mu}_B p^{\nu}_B F^{B \rightarrow
T}(m_P^2)) (a_3 -a_5)~,  \nonumber \\
P^{(\prime)}_T &=& -i {G_F \over \sqrt{2}} V^*_{tb} V_{td (s)}
(f_P \epsilon^*_{\mu \nu} p^{\mu}_B p^{\nu}_B F^{B \rightarrow
T}(m_P^2)) (a_4 -2 a_6 X_{q q^{\prime}})~,  \nonumber \\
P_{EW,T}^{(\prime)} &=& -i {G_F \over \sqrt{2}} V^*_{tb} V_{td
(s)} (f_P \epsilon^*_{\mu \nu} p^{\mu}_B p^{\nu}_B F^{B
\rightarrow T}(m_P^2)) {3 \over 2} (a_7 -a_9)~, \nonumber \\
P_{EW,T}^{C (\prime)} &=& -i {G_F \over \sqrt{2}} V^*_{tb} V_{td
(s)} (f_P \epsilon^*_{\mu \nu} p^{\mu}_B p^{\nu}_B F^{B
\rightarrow T}(m_P^2)) {3 \over 2} (a_{10} -2 a_8 X_{q
q^{\prime}})~, \label{su3facto}
\end{eqnarray}
where
\begin{eqnarray}
F^{B \rightarrow T}(m_P^2) &=& k(m_P^2) +(m_B^2 -m_T^2) b_+(m_P^2)
+m_P^2 b_-(m_P^2)~, \label{FBT} \\
X_{q q^{\prime}} &=& {m_P^2 \over (m_b +m_{q^{\prime}}) (m_q
+m_{q^{\prime}})}~.  \label{Xqq}
\end{eqnarray}
Here the effective coefficients $a_i$ are defined as $a_i =
c^{eff}_i + \xi c^{eff}_{i+1}$ ($i =$ odd) and $a_i = c^{eff}_i +
\xi c^{eff}_{i-1}$ ($i =$ even) with the effective WC's
$c^{eff}_i$ at the scale $m_b$ \cite{10}, and by treating
$\xi\equiv 1/N_c$ ($N_c$ denotes the effective number of color) as
an adjustable parameter.  The last term with $b_-$ in Eq.
(\ref{FBT}) gives a negligible contribution to the decay amplitude
due to the small mass factor.

With Tables I, II and the above relations (\ref{su3facto}), one
can easily write down in the factorization scheme the amplitude of
any $B \rightarrow PT$ mode shown in the tables.  For example,
from Table I and the relations (\ref{su3facto}), the amplitude of
the process $B^+ \rightarrow \pi^+ a_2^0$ can be written
as\footnote{In the factorization scheme, we use the usual phase
convention for the pseudoscalar and the tensor mesons as follows:
$\pi^0 (a^0_2) = {1 \over \sqrt{2}} (u \bar u -d \bar d)$,  $\pi^-
(a^-_2) = \bar u d$,  $K^- (K^{*-}_2) = \bar u s$. }
\begin{eqnarray}
A (B^+ \rightarrow \pi^+ a^0_2) &=& -{1 \over \sqrt{2}} \left(T_T
+ C_P +P_T -P_P +P_{EW,P} +{2 \over 3} P^C_{EW,T} +{1 \over 3}
P^C_{EW,P} \right) \nonumber \\
&=& i {G_F \over 2} f_{\pi} \epsilon^*_{\mu \nu} p^{\mu}_B
p^{\nu}_B F^{B \rightarrow a^0_2} (m^2_{\pi}) \left\{ V^*_{ub}
V_{ud} a_1 - V^*_{tb} V_{td} [a_4 +a_{10} -2 (a_6 +a_8) X_{du} ]
\right\} .
\end{eqnarray}
Here we have used the fact that $C_P$, $P_P$, $P_{EW,P}$, and
$P^C_{EW,P}$ with the subscript $P$ all vanish.
In Appendix, expressions for
all the amplitudes of $B \rightarrow PT$ decays 
are presented as calculated in the factorization scheme.

To calculate the unpolarized decay rates for $B \rightarrow PT$,
we sum over polarizations of the tensor meson $T$ using the
following formula \cite{2}:
\begin{eqnarray}
\sum_{\lambda} \epsilon_{\alpha \beta}(p_{_T}, \lambda)
\epsilon^*_{\mu \nu}(p_{_T}, \lambda) = {1 \over 2}
(\theta_{\alpha \mu} \theta_{\beta \nu} +\theta_{\beta \mu}
\theta_{\alpha \nu}) -{1 \over 3} \theta_{\alpha \beta}
\theta_{\mu \nu}~,
\end{eqnarray}
where $\theta_{\alpha \beta} = - g_{\alpha \beta}
+(p_{_T})_{\alpha} (p_{_T})_{\beta}/ m_T^2$.  Then, the decay rate
for $B \rightarrow PT$ is given by
\begin{eqnarray}
\Gamma (B \rightarrow PT) = { |\vec{p}_{_P}|^5 \over 12 \pi m_T^2}
\left( {m_B \over m_T} \right)^2 \left| {A (B \rightarrow PT)
\over \epsilon^*_{\mu \nu} p^{\mu}_B p^{\nu}_B} \right|^2 ,
\end{eqnarray}
where $|\vec{p}_{_P}|$ is the magnitude of three-momentum of the
final state particle $P$ or $T$ ($|\vec{p}_{_P}|=|\vec{p}_{_T}|$)
in the rest frame of the $B$ meson.  The CP asymmetry, ${\mathcal
A}_{CP}$, is defined by
\begin{eqnarray}
{\mathcal A}_{CP} = {{\mathcal B}(b \rightarrow f)  -{\mathcal
B}(\bar b \rightarrow \bar f) \over {\mathcal B}(b \rightarrow f)
+{\mathcal B}(\bar b \rightarrow \bar f)}~,
\end{eqnarray}
where $b$ and $f$ denote $b$ quark and a generic final state,
respectively.

We calculate the branching ratios and CP asymmetries for $B
\rightarrow PT$ decay modes for various input parameter values.
The predictions are sensitive to several input parameters such as
the form factors, the strange quark mass, the parameter $\xi
\equiv 1/ N_c$, the CKM
matrix elements and in particular, the weak phase $\gamma$. In a
recent work \cite{10} on charmless $B$ decays to two light mesons
such as $PP$ and $VP$, it has been shown that the favored values
of the input parameters are
$$
\xi \approx 0.45,~ m_s(m_b) \approx 85 ~{\rm MeV},~\gamma \approx
110^0,~ V_{cb} =0.040,~~~{\rm and}~~~ |V_{ub} /V_{cb}| =0.087
$$
in order to get the best fit to the recent
experimental data from the CLEO collaboration.  For our numerical
calculations, we use the following values of the decay constants
(in MeV units) \cite{9,13,14}:
$$
f_{\pi} =132,~~ f_{\eta} =131,~~
f_{\eta^{\prime}} =118,~~f_{K} =162.
$$
We use the values of the
form factors for the $B \rightarrow T$ transition calculated in
the ISGW model \cite{4}. The strange quark mass $m_s$ is in
considerable doubt: {\it i.e.}, QCD sum rules give $m_s(1\; {\rm
GeV})=(175\pm 25)$ MeV and lattice gauge theory gives $m_s(2\;
{\rm GeV})=(100 \pm 20 \pm 10)$ MeV in the quenched lattice
calculation \cite{15}. In this analysis we use two representative
values of $m_s = 100$ MeV and $m_s = 85$ MeV at $m_b$ scale.
Current best estimates for CKM matrix elements are $V_{cb} =0.0381
\pm 0.0021$ and $|V_{ub} /V_{cb}| =0.085 \pm 0.019$ \cite{16}.  We
use $V_{cb} =0.040$ and $|V_{ub} /V_{cb}| =0.087$.  It has been
known that there exists the discrepancy in values of $\gamma$
extracted from CKM-fitting at $\rho-\eta$ plane \cite{17} and from
the $\chi^2$ analysis of non-leptonic decays of $B$ mesons
\cite{18,19}.  The value of $\gamma$ obtained from unitarity
triangle fitting is in the range of $60^0 \sim 80^0$. But in
analysis of non-leptonic $B$ decay, possibility of larger $\gamma$
has been discussed by Deshpande {\it et al.} \cite{18} and He {\it
et al.} \cite{19}. The obtained value of $\gamma$ is $\gamma =90^0
\sim 140^0$.  In our calculations we use two representative values
of $\gamma = 110^0$ and $\gamma =65^0$.

In Tables III $-$ VI, we show the branching ratios and the CP
asymmetries for $B \rightarrow PT$ decays with either $\Delta S
=0$ or $|\Delta S| =1$.  In the tables the second and the third
columns correspond to the sets of the input parameters,
$$
\{\xi =0.1,~ m_s = 85~{\rm MeV},~ \gamma =110^0 \}~~~ {\rm and}~~~
\{\xi =0.1,~ m_s = 100~{\rm MeV},~ \gamma =65^0\}~,
$$
respectively. Similarly, the fourth
and the fifth columns correspond to the cases,
$$
\{\xi =0.3,~ m_s = 85~{\rm MeV},~ \gamma =110^0\}~~~{\rm and}~~~
\{\xi =0.3,~ m_s = 100~{\rm MeV},~ \gamma =65^0\}~,
$$
respectively.  The sixth and the seventh columns
correspond to the cases,
$$
\{\xi =0.5,~ m_s = 85~{\rm MeV},~ \gamma =110^0\}~~~{\rm and}~~~
\{\xi =0.5,~ m_s = 100~{\rm MeV}~, \gamma =65^0\}~,
$$
respectively.  Here $\xi \equiv 1/ N_c = 0.3$ corresponds to the
case of naive factorization ($N_c =3$).  It has been known that in
$B \rightarrow D$ decays the generalized factorization has been
successfully used with the favored value of $\xi \approx 0.5$
\cite{20}.  Also, as mentioned above, a recent analysis of
charmless $B$ decays to two light mesons such as $PP$ and $VP$
\cite{10} shows that $\xi \approx 0.45$ is favored with certain
values of other parameters for the best fit to the recent CLEO
data.

The branching ratios and the CP asymmetries for $B \rightarrow PT$
decay modes with $\Delta S =0$ are shown in Table III and IV.
Among $\Delta S=0$ modes, the decay modes $B^+ \rightarrow \pi^+
a^0_2$, $B^+ \rightarrow \pi^+ f_2$, and $B^0 \rightarrow \pi^+
a^-_2$ have relatively large branching ratios of a few times
$10^{-7}$.  This prediction is consistent with that based on
flavor SU(3) symmetry.  We see that in the factorization scheme
the following equality between the branching ratios holds for any
set of the parameters given above: $2 {\cal B}(B^+ \rightarrow
\pi^+ a^0_2) \approx {\cal B}(B^0 \rightarrow \pi^+ a^-_2)$, as
discussed in Eq. (\ref{facto1}).  (Little deviation from the exact
equality arises from breaking of isospin symmetry.) We also see
from Table III that ${\cal B}(B^+ \rightarrow \pi^0 a^+_2)$ is
much smaller than ${\cal B}(B^+ \rightarrow \pi^+ a^0_2)$ by an
order of magnitude or even three orders of magnitude depending on
values of the input parameters, because in factorization the
dominant contribution to the former mode arises from the
color-suppressed tree diagram ($C_T$), while the dominant one to
the latter mode arises from the color-favored tree diagram
($T_T$).  Note that in flavor SU(3) symmetry the amplitude for
$B^+ \rightarrow \pi^+ a^0_2$ has the color-favored tree
contribution $T_P$ constructive to the color-suppressed tree
contribution $C_T$ (in addition to small contributions from the
penguin diagrams). (Also recall that the magnitude of $T_P$ can be
possibly measured by Eq. (\ref{b0pia2})).  In case that $T_P$ is
not small compared to $T_T$, ${\cal B}(B^+ \rightarrow \pi^0
a^+_2)$ can be comparable to ${\cal B}(B^+ \rightarrow \pi^+
a^0_2)$.  Therefore, measurement of the modes $B \rightarrow \pi
a_2$ in future experiment will provide important information on
the above discussion. Some $\Delta S=0$ processes such as $B^+
\rightarrow \eta^{\prime} a^+_2$, $B^0 \rightarrow \eta^{\prime}
a^0_2$, and $B^0 \rightarrow \eta^{\prime} f_2$ have the branching
ratios of order of $10^{-7}$.  The branching ratios of the other
processes are order of $10^{-8}$ or less.  The CP asymmetry for
$B^+ \rightarrow \eta^{\prime} a^+_2$ is relatively large (about
$20 \%$ or larger) with the branching ratio of order of $10^{-7}$
for $\xi =0.5$ and $0.1$.  The CP asymmetry for $B^+ \rightarrow
\eta a^+_2$, $B^0 \rightarrow \eta a^0_2$, $\eta f_2$ can be as
large as $71 \%$ for $\xi =0.5$, with the branching ratios of
$O(10^{-8})$.

Tables V shows the branching ratios for $|\Delta S|=1$ decay
processes. In $|\Delta S|=1$ decays, the relevant penguin diagrams
give dominant contribution to the decay rates. We see that the
branching ratios for $|\Delta S|=1$ decays are in range between
$O(10^{-7})$ and $O(10^{-10})$, similar to those for $\Delta S =0$
decays.  The modes $B^{+ (0)} \rightarrow \eta^{\prime} K_2^{*
+(0)}$ have relatively larger branching ratios of $O(10^{-7})$. In
contrast, the modes $B^{+ (0)} \rightarrow \eta K_2^{* +(0)}$ have
very small branching ratios of $O(10^{-9})$ to $O(10^{-10})$.
Based on flavor SU(3) symmetry, this fact has been expected by
observation that the penguin contributions $P^{\prime}$ and
$S^{\prime}$ interfere constructively for $B^{+ (0)} \rightarrow
\eta^{\prime} K_2^{* +(0)}$, but destructively for $B^{+ (0)}
\rightarrow \eta K_2^{* +(0)}$.  From Eq. (\ref{su3facto}), we see
that $P_T^{\prime}$ and $S_T^{\prime}$ are proportional to $(a_4
-2 a_6 X_{ss})$ and $(a_3 -a_5)$, respectively, in addition to
other common factors. Indeed, in the factorization scheme, since
$(a_4 -2 a_6 X_{ss})$ and $(a_3 -a_5)$ have the same sign (all
positive), the combination $(2 P_T^{\prime} +4 S_T^{\prime})$
appearing in $B^{+ (0)} \rightarrow \eta^{\prime} K_2^{* +(0)}$
causes constructive interference, while the combination $(-
P_T^{\prime} + S_T^{\prime})$ appearing in $B^{+ (0)} \rightarrow
\eta K_2^{* +(0)}$ causes destructive interference (see Appendix).
Thus the prediction for these decay modes are consistent in both
approaches.  The modes $B^+ \rightarrow \pi^0 K_2^{* +}$ and $B^0
\rightarrow \pi^0 K_2^{* 0}$ have almost the same branching ratios
of $O(10^{-8})$ in the factorization scheme (also see Appendix).
In flavor SU(3) symmetry, as shown in Table II, the decay
amplitudes for these modes have the contributions from
$P_P^{\prime}$ or $T_P^{\prime}$.  Thus, as discussed in the
previous section, if $P_P^{\prime}$ or $T_P^{\prime}$ is not very
suppressed compared to $C_T^{\prime}$, then there would be a
sizable discrepancy in ${\cal B} (B^+ \rightarrow \pi^0 K_2^{*+})
\approx {\cal B} (B^0 \rightarrow \pi^0 K_2^{*0})$, and in
principle it can be tested in experiment.  The terms
$-(T_P^{\prime} +P_P^{\prime} +{2 \over 3} P_{EW,P}^{C \prime})$
and $P_P^{\prime} -{1 \over 3} P_{EW,P}^{C \prime}$ can be
determined by measuring the branching ratios for $B^0 \rightarrow
\pi^- K_2^{*+}$ and $B^+ \rightarrow \pi^+ K_2^{*0}$,
respectively. The CP asymmetries ${\cal A_{CP}}$ in $|\Delta S|=1$
decays are shown in Table VI. ${\cal A_{CP}}$'s in most modes are
expected to be quite small. In $B^+(0) \rightarrow \eta
K^{*+(0)}_2$, ${\cal A_{CP}}$ can be as large as $92 \%$, but the
corresponding branching ratio is as small as about $O(10^{-9})$.

\section{conclusion}

We have analyzed exclusive charmless decays $B \rightarrow PT$ in
the schemes of both flavor SU(3) symmetry and generalized
factorization.  Using the flavor SU(3) symmetry, we have
decomposed all the amplitudes for decays $B \rightarrow PT$ into
linear combinations of the relevant SU(3) amplitudes. Based on the
decomposition, we have shown that certain decay modes, such as
$B^+ \rightarrow \pi^+ a_2^0$, $\pi^+ f_2$ and $B^0 \rightarrow
\pi^+ a_2^-$ in $\Delta S =0$ decays, and $B^{+(0)} \rightarrow
\eta^{\prime} K^{*+(0)}_2$ in $|\Delta S| =1$ decays, are expected
to have the largest decay rates and so these modes can be
preferable to find in future experiment. Certain ways to test
validity of the factorization scheme have been presented by
emphasizing interplay between both approaches and carefully
combining the predictions from both approaches.  In order to
bridge the flavor SU(3) approach and the factorization approach,
we have explicitly presented a set of relations between a flavor
SU(3) amplitude and a corresponding amplitude in factorization in
$B \rightarrow PT$ decays.

To calculate the branching ratios for $B \rightarrow PT$ decays,
we have used the \emph{full} effective Hamiltonian including all
the penguin operators which are essential to analyze the $|\Delta
S| =1$ processes and to calculate CP asymmetries.  We have also
used the non-relativistic quark model proposed by Isgur, Scora,
Grinstein, and Wise to obtain the form factors describing $B
\rightarrow T$ transitions. As shown in Tables III and V, the
branching ratios vary from $O(10^{-7})$ to $O(10^{-10})$.
Consistent with the prediction from the flavor SU(3) analysis, the
decay modes such as $B^+ \rightarrow \pi^+ a_2^0$, $\pi^+ f_2$,
$B^0 \rightarrow \pi^+ a_2^-$ and $B^{+(0)} \rightarrow
\eta^{\prime} K^{*+(0)}_2$ as well as $B^+ \rightarrow
\eta^{\prime} a_2^+$ have the branching ratios of order of
$10^{-7}$.  In particular, the branching ratio for the mode $B^0
\rightarrow \pi^+ a_2^-$ can be as large as almost $O(10^{-6})$.
We have identified the decay modes where the CP asymmetries are
expected to be large, such as $B \rightarrow \eta^{\prime} a^+_2$,
$\eta a^+_2$, $\eta a^0_2$, $\eta f_2$ in $\Delta S =0$ decays,
and $B^+(0) \rightarrow \eta K^{*+(0)}_2$ in $|\Delta S| =1$
decays. Due to possible uncertainties in the hadronic form factors
of $B \to PT$ and non-factorizaton effects, the predicted
branching ratios could be increased by even two orders of
magnitude for some decay modes \cite{newwork}. Although
experimentally challenging, the exclusive charmless decays, $B \to
PT$, can probably be carried out in details at hadronic $B$
experiments such as BTeV and LHC-B, where more than $10^{10}$
$B$-mesons will be produced per year, as well as at present
asymmetric $B$ factories of Belle and Babar.
\\

\centerline{\bf ACKNOWLEDGEMENTS}
\medskip

\noindent The work of C.S.K. was supported in part by  Grant No.
2000-1-11100-003-1 and SRC Program of the KOSEF, and in part by
the KRF Grants, Project No. 2000-015-DP0077. The work of B.H.L and
S.O. were supported by the KRF Grants, £ Project No.
2000-015-DP0077 and by the BK21 Program.


\newpage
\begin{center}
{\bf APPENDIX}
\end{center}

In this Appendix, we present expressions for all the decay
amplitudes of $B \rightarrow PT$ modes shown in Tables I and II as
calculated in the factorization scheme.  Below we use $F^{B \to
T}$ and $X_{q q^{\prime}}$ defined in Eqs. (\ref{FBT}) and
(\ref{Xqq}).
\\
(1) $B \rightarrow PT$ ($\Delta S = 0$) decays.
\begin{eqnarray}
A(B^+ \rightarrow \pi^+ a^0_2) &=& i {G_F \over 2} f_{\pi}
\epsilon^*_{\mu \nu} p^{\mu}_B p^{\nu}_B F^{B \rightarrow a^0_2}
(m^2_{\pi}) \left\{ V^*_{ub} V_{ud} a_1 - V^*_{tb} V_{td} [a_4
+a_{10} -2 (a_6 +a_8) X_{du} ] \right\},
\\
A(B^+ \rightarrow \pi^+ f_2) &=& i {G_F \over 2} \cos\phi_{_T}
f_{\pi} \epsilon^*_{\mu \nu} p^{\mu}_B p^{\nu}_B F^{B \rightarrow
f_2} (m^2_{\pi}) \left\{ V^*_{ub} V_{ud} a_1 \right.
\nonumber \\
&\mbox{}& \left. - V^*_{tb} V_{td} [a_4 +a_{10} -2 (a_6 +a_8)
X_{du} ] \right\}
\\
A(B^+ \rightarrow \pi^+ f^{\prime}_2) &=& i {G_F \over 2}
\sin\phi_{_T} f_{\pi} \epsilon^*_{\mu \nu} p^{\mu}_B p^{\nu}_B
F^{B \rightarrow f_2} (m^2_{\pi}) \left\{ V^*_{ub} V_{ud} a_1
\right.
\nonumber \\
&\mbox{}& \left. - V^*_{tb} V_{td} [a_4 +a_{10} -2 (a_6 +a_8)
X_{du} ] \right\}
\\
A(B^+ \rightarrow \pi^0 a^+_2) &=& i {G_F \over 2} f_{\pi}
\epsilon^*_{\mu \nu} p^{\mu}_B p^{\nu}_B F^{B
\rightarrow a^+_2} (m^2_{\pi}) \left\{ V^*_{ub} V_{ud} a_2 \right.
\nonumber \\
&\mbox{}& \left. - V^*_{tb} V_{td} \left[-a_4 +{3 \over 2} a_7 -{3
\over 2} a_9 +{1 \over 2} a_{10} +2 \left(a_6 -{1 \over 2} a_8
\right) X_{dd} \right] \right\}
\\
A(B^+ \rightarrow \eta a^+_2) &=& i {G_F \over \sqrt{2}} {1 \over
\sqrt{3}} f_{\eta} \epsilon^*_{\mu \nu}
p^{\mu}_B p^{\nu}_B F^{B \rightarrow a^+_2} (m^2_{\eta}) \left\{
V^*_{ub} V_{ud} a_2 \right.
\nonumber   \\
&\mbox{}& \left. - V^*_{tb} V_{td} \left[a_3 +a_4 -a_5 +a_7 -a_9
-{1 \over 2} a_{10} -2 \left(a_6 -{1 \over 2} a_8 \right) X_{dd}
\right] \right\}
\\
A(B^+ \rightarrow \eta^{\prime} a^+_2) &=& i {G_F \over \sqrt{2}}
{1 \over \sqrt{6}} f_{\eta^{\prime}} \epsilon^*_{\mu \nu}
p^{\mu}_B p^{\nu}_B F^{B \rightarrow a^+_2} (m^2_{\eta^{\prime}})
\left\{ V^*_{ub} V_{ud} a_2 \right. \nonumber \\ &\mbox{}& \left.
- V^*_{tb} V_{td} \left[4 a_3 +a_4 -4 a_5 -{1 \over 2} a_7 +{1
\over 2} a_9 -{1 \over 2} a_{10} -2 \left(a_6 -{1 \over 2} a_8
\right) X_{dd} \right] \right\}
\\
A(B^+ \rightarrow \bar K^0 K^{*+}_2) &=& -i {G_F \over \sqrt{2}}
f_{K} \epsilon^*_{\mu \nu} p^{\mu}_B p^{\nu}_B F^{B \rightarrow
K^{*+}_2} (m^2_{K}) V^*_{tb} V_{td} \left[a_4 -{1 \over 2} a_{10}
-2 \left(a_6 -{1 \over 2} a_8 \right) X_{ds} \right]
\\
A(B^0 \rightarrow \pi^+ a^-_2) &=& i {G_F \over \sqrt{2}} f_{\pi}
\epsilon^*_{\mu \nu} p^{\mu}_B p^{\nu}_B F^{B \rightarrow a^-_2}
(m^2_{\pi}) \left\{ V^*_{ub} V_{ud} a_1 - V^*_{tb} V_{td} [a_4
+a_{10} -2 (a_6 +a_8) X_{du} ] \right\}
\\
A(B^0 \rightarrow \pi^0 a^0_2) &=& i {G_F \over 2\sqrt{2}} f_{\pi}
\epsilon^*_{\mu \nu} p^{\mu}_B p^{\nu}_B
F^{B \rightarrow a^0_2} (m^2_{\pi}) \left\{ V^*_{ub} V_{ud} (-a_2)
\right.
\nonumber \\
&\mbox{}& \left. - V^*_{tb} V_{td} \left[a_4 -{3 \over 2} a_7 +{3
\over 2} a_9 -{1 \over 2} a_{10} -2 \left(a_6 -{1 \over 2} a_8
\right) X_{dd} \right] \right\}
\\
A(B^0 \rightarrow \pi^0 f_2) &=& i {G_F \over 2\sqrt{2}}
\cos\phi_{_T} f_{\pi} \epsilon^*_{\mu \nu} p^{\mu}_B p^{\nu}_B
F^{B \rightarrow f_2} (m^2_{\pi}) \left\{ V^*_{ub} V_{ud} (-a_2)
\right.
\nonumber \\
&\mbox{}& \left. - V^*_{tb} V_{td} \left[a_4 -{3 \over 2} a_7 +{3
\over 2} a_9 -{1 \over 2} a_{10} -2 \left(a_6 -{1 \over 2} a_8
\right) X_{dd} \right] \right\}
\\
A(B^0 \rightarrow \pi^0 f^{\prime}_2) &=& i {G_F \over 2\sqrt{2}}
\sin\phi_{_T} f_{\pi} \epsilon^*_{\mu \nu} p^{\mu}_B p^{\nu}_B
F^{B \rightarrow f^{\prime}_2} (m^2_{\pi}) \left\{ V^*_{ub} V_{ud}
(-a_2) \right.
\nonumber \\
&\mbox{}& \left. - V^*_{tb} V_{td} \left[a_4 -{3 \over 2} a_7 +{3
\over 2} a_9 -{1 \over 2} a_{10} -2 \left(a_6 -{1 \over 2} a_8
\right) X_{dd} \right] \right\}
\\
A(B^0 \rightarrow \eta a^0_2) &=& i {G_F \over \sqrt{2}} {1 \over
\sqrt{6}} f_{\eta} \epsilon^*_{\mu \nu}
p^{\mu}_B p^{\nu}_B F^{B \rightarrow a^0_2} (m^2_{\eta}) \left\{
V^*_{ub} V_{ud} a_2 \right.
\nonumber   \\
&\mbox{}& \left. - V^*_{tb} V_{td} \left[a_3 +a_4 -a_5 +a_7 -a_9
-{1 \over 2} a_{10} -2 \left(a_6 -{1 \over 2} a_8 \right) X_{dd}
\right] \right\}
\\
A(B^0 \rightarrow \eta f_2) &=& i {G_F \over \sqrt{2}} {1 \over
\sqrt{6}} \cos\phi_{_T} f_{\eta} \epsilon^*_{\mu \nu} p^{\mu}_B
p^{\nu}_B F^{B \rightarrow f_2} (m^2_{\eta}) \left\{
V^*_{ub} V_{ud} a_2 \right.
\nonumber \\
&\mbox{}& \left. - V^*_{tb} V_{td} \left[a_3 +a_4 -a_5 +a_7 -a_9
-{1 \over 2} a_{10} -2 \left(a_6 -{1 \over 2} a_8 \right) X_{dd}
\right] \right\}
\\
A(B^0 \rightarrow \eta f^{\prime}_2) &=& i {G_F \over \sqrt{2}} {1
\over \sqrt{6}} \sin\phi_{_T} f_{\eta} \epsilon^*_{\mu \nu}
p^{\mu}_B p^{\nu}_B F^{B \rightarrow f^{\prime}_2} (m^2_{\eta})
\left\{ V^*_{ub} V_{ud} a_2 \right.
\nonumber   \\
&\mbox{}& \left. - V^*_{tb} V_{td} \left[a_3 +a_4 -a_5 +a_7 -a_9
-{1 \over 2} a_{10} -2 \left(a_6 -{1 \over 2} a_8 \right) X_{dd}
\right] \right\}
\\
A(B^0 \rightarrow \eta^{\prime} a^0_2) &=& i {G_F \over \sqrt{2}}
{1 \over 2\sqrt{3}} f_{\eta^{\prime}} \epsilon^*_{\mu \nu}
p^{\mu}_B p^{\nu}_B F^{B \rightarrow a^0_2} (m^2_{\eta^{\prime}})
\left\{ V^*_{ub} V_{ud} a_2
\right.
\nonumber \\
&\mbox{}& \left. - V^*_{tb} V_{td} \left[4 a_3 +a_4 -4 a_5 -{1
\over 2} a_7 +{1 \over 2} a_9 -{1 \over 2} a_{10} -2 \left(a_6 -{1
\over 2} a_8 \right) X_{dd} \right] \right\}
\\
A(B^0 \rightarrow \eta^{\prime} f_2) &=& i {G_F \over \sqrt{2}} {1
\over 2\sqrt{3}} \cos\phi_{_T} f_{\eta^{\prime}} \epsilon^*_{\mu
\nu} p^{\mu}_B p^{\nu}_B F^{B \rightarrow f_2}
(m^2_{\eta^{\prime}}) \left\{ V^*_{ub} V_{ud} a_2 \right.
\nonumber \\
&\mbox{}& \left. - V^*_{tb} V_{td} \left[4 a_3 +a_4 -4 a_5 -{1
\over 2} a_7 +{1 \over 2} a_9 -{1 \over 2} a_{10} -2 \left(a_6 -{1
\over 2} a_8 \right) X_{dd} \right] \right\}
\\
A(B^0 \rightarrow \eta^{\prime} f^{\prime}_2) &=& i {G_F \over
\sqrt{2}} {1 \over 2\sqrt{3}} \sin\phi_{_T} f_{\eta^{\prime}}
\epsilon^*_{\mu \nu} p^{\mu}_B p^{\nu}_B F^{B \rightarrow f_2}
(m^2_{\eta^{\prime}}) \left\{ V^*_{ub} V_{ud} a_2 \right.
\nonumber \\
&\mbox{}& \left. - V^*_{tb} V_{td} \left[4 a_3 +a_4 -4 a_5 -{1
\over 2} a_7 +{1 \over 2} a_9 -{1 \over 2} a_{10} -2 \left(a_6 -{1
\over 2} a_8 \right) X_{dd} \right] \right\}
\\
A(B^0 \rightarrow \bar K^0 K^{*0}_2) &=& -i {G_F \over \sqrt{2}}
f_{K} \epsilon^*_{\mu \nu} p^{\mu}_B p^{\nu}_B F^{B \rightarrow
K^{*0}_2} (m^2_{K}) V^*_{tb} V_{td} \left[a_4 -{1 \over 2} a_{10}
-2 \left(a_6 -{1 \over 2} a_8 \right) X_{ds} \right]
\end{eqnarray}  \\
(2) $B \rightarrow PT$ ($|\Delta S| = 1$) decays.
\begin{eqnarray}
A(B^+ \rightarrow K^+ a^0_2) &=& i {G_F \over 2} f_{K}
\epsilon^*_{\mu \nu} p^{\mu}_B p^{\nu}_B F^{B \rightarrow a^0_2}
(m^2_{K}) \left\{ V^*_{ub} V_{us} a_1 - V^*_{tb} V_{ts} [a_4
+a_{10} -2 (a_6 +a_8) X_{su} ] \right\}
\\
A(B^+ \rightarrow K^+ f_2) &=& i {G_F \over 2} \cos\phi_{_T} f_{K}
\epsilon^*_{\mu \nu} p^{\mu}_B p^{\nu}_B F^{B \rightarrow f_2}
(m^2_{K}) \left\{ V^*_{ub} V_{us} a_1 \right.
\nonumber \\
&\mbox{}& \left. - V^*_{tb} V_{ts} [a_4 +a_{10} -2 (a_6 +a_8)
X_{su} ] \right\}
\\
A(B^+ \rightarrow K^+ f^{\prime}_2) &=& i {G_F \over 2}
\sin\phi_{_T} f_{K} \epsilon^*_{\mu \nu} p^{\mu}_B p^{\nu}_B F^{B
\rightarrow f^{\prime}_2} (m^2_{K}) \left\{ V^*_{ub} V_{us} a_1
\right.
\nonumber \\
&\mbox{}& \left. - V^*_{tb} V_{ts} [a_4 +a_{10} -2 (a_6 +a_8)
X_{su} ] \right\}
\\
A(B^+ \rightarrow \bar K^0 a^+_2) &=& -i {G_F \over \sqrt{2}}
f_{K} \epsilon^*_{\mu \nu} p^{\mu}_B p^{\nu}_B F^{B \rightarrow
a^+_2} (m^2_{K}) V^*_{tb} V_{ts} \left[a_4 -{1 \over 2} a_{10} -2
\left(a_6 -{1 \over 2} a_8 \right) X_{sd} \right]
\\
A(B^+ \rightarrow \pi^0 K^{*+}_2) &=& i {G_F \over 2} f_{\pi}
\epsilon^*_{\mu \nu} p^{\mu}_B p^{\nu}_B F^{B \rightarrow
K^{*+}_2} (m^2_{\pi}) \left[ V^*_{ub} V_{us} a_2 - V^*_{tb} V_{ts}
\left( {3 \over 2} a_7 -{3 \over 2} a_9 \right) \right]
\\
A(B^+ \rightarrow \eta K^{*+}_2) &=& i {G_F \over \sqrt{2}} {1
\over \sqrt{3}} f_{\eta} \epsilon^*_{\mu \nu}
p^{\mu}_B p^{\nu}_B F^{B \rightarrow K^{*+}_2} (m^2_{\eta}) \left\{
V^*_{ub} V_{us} a_2 \right.
\nonumber \\
&\mbox{}& \left. - V^*_{tb} V_{ts} \left[a_3 -a_4 -a_5 +a_7 -a_9
+{1 \over 2} a_{10} +2 \left(a_6 -{1 \over 2} a_8 \right) X_{ss}
\right] \right\}
\\
A(B^+ \rightarrow \eta^{\prime} K^{*+}_2) &=& i {G_F \over \sqrt{2}}
{1 \over \sqrt{6}} f_{\eta^{\prime}}
\epsilon^*_{\mu \nu} p^{\mu}_B p^{\nu}_B F^{B \rightarrow
K^{*+}_2} (m^2_{\eta^{\prime}}) \left\{ V^*_{ub} V_{us}
a_2 \right.
\nonumber \\
&\mbox{}& \left. - V^*_{tb} V_{ts}\left[4 a_3 +2 a_4 -4 a_5 -{1
\over 2} a_7 +{1 \over 2} a_9 - a_{10} -4 \left(a_6 -{1 \over 2}
a_8 \right) X_{ss} \right] \right\}
\\
A(B^0 \rightarrow K^+ a^-_2) &=& i {G_F \over \sqrt{2}} f_{K}
\epsilon^*_{\mu \nu} p^{\mu}_B p^{\nu}_B F^{B \rightarrow a^-_2}
(m^2_{K}) \left\{ V^*_{ub} V_{us} a_1 - V^*_{tb} V_{ts} [a_4
+a_{10} -2 (a_6 +a_8) X_{su} ] \right\}
\\
A(B^0 \rightarrow K^0 a^0_2) &=& i {G_F \over 2} f_{K}
\epsilon^*_{\mu \nu} p^{\mu}_B p^{\nu}_B F^{B \rightarrow a^0_2}
(m^2_{K}) V^*_{tb} V_{ts} \left[a_4 -{1 \over 2} a_{10} -2
\left(a_6 -{1 \over 2} a_8 \right) X_{sd} \right]
\\
A(B^0 \rightarrow K^0 f_2) &=& i {G_F \over 2} \cos\phi_{_T} f_{K}
\epsilon^*_{\mu \nu} p^{\mu}_B p^{\nu}_B F^{B \rightarrow f_2}
(m^2_{K}) V^*_{tb} V_{ts} \left[a_4 -{1 \over 2} a_{10} -2
\left(a_6 -{1 \over 2} a_8 \right) X_{sd} \right]
\\
A(B^0 \rightarrow K^0 f^{\prime}_2) &=& i {G_F \over 2}
\sin\phi_{_T} f_{K} \epsilon^*_{\mu \nu} p^{\mu}_B p^{\nu}_B F^{B
\rightarrow f^{\prime}_2} (m^2_{K}) V^*_{tb} V_{ts} \left[a_4 -{1
\over 2} a_{10} -2 \left(a_6 -{1 \over 2} a_8 \right) X_{sd}
\right]
\\
A(B^0 \rightarrow \pi^0 K^{*0}_2) &=& i {G_F \over 2} f_{\pi}
\epsilon^*_{\mu \nu} p^{\mu}_B p^{\nu}_B F^{B \rightarrow
K^{*0}_2} (m^2_{\pi}) \left[ V^*_{ub} V_{us} a_2 - V^*_{tb} V_{ts}
\left( {3 \over 2} a_7 -{3 \over 2} a_9 \right) \right]
\\
A(B^0 \rightarrow \eta K^{*0}_2) &=& i {G_F \over \sqrt{2}} {1
\over \sqrt{3}} f_{\eta} \epsilon^*_{\mu \nu}
p^{\mu}_B p^{\nu}_B F^{B \rightarrow K^{*0}_2} (m^2_{\eta})
\left\{ V^*_{ub} V_{us} a_2 \right.
\nonumber \\
&\mbox{}& \left. - V^*_{tb} V_{ts} \left[a_3 -a_4 -a_5 +a_7 -a_9
+{1 \over 2} a_{10} +2 \left(a_6 -{1 \over 2} a_8 \right) X_{ss}
\right] \right\}
\\
A(B^0 \rightarrow \eta^{\prime} K^{*0}_2) &=& i {G_F \over
\sqrt{2}} {1 \over \sqrt{6}} f_{\eta^{\prime}} \epsilon^*_{\mu
\nu} p^{\mu}_B p^{\nu}_B F^{B \rightarrow K^{*0}_2}
(m^2_{\eta^{\prime}}) \left\{ V^*_{ub} V_{us}
a_2 \right.
\nonumber \\
&\mbox{}& \left. - V^*_{tb} V_{ts}\left[4 a_3 +2 a_4 -4 a_5 -{1
\over 2} a_7 +{1 \over 2} a_9 - a_{10} -4 \left(a_6 -{1 \over 2}
a_8 \right) X_{ss} \right] \right\}
\end{eqnarray}

\newpage
\begin{table}
\caption{Coefficients of SU(3) amplitudes in $B \rightarrow PT$
($\Delta S = 0$). The coefficients of the SU(3) amplitudes with
the subscript $P$ are expressed in square brackets.  As explained
in Sec. II, the contributions of the SU(3) amplitudes with the
subscript $P$ vanish in the framework of factorization, because
those contributions contain the matrix element $\langle T|
J^{\rm{weak}}_{\mu} |0 \rangle$, which is zero.  Here $c$ and $s$
denote $\cos \phi_{_T}$ and $\sin \phi_{_T}$, respectively. }
\begin{tabular}{c||c|c|c|c|c|c|c}
$B \rightarrow PT$ & factor & $T_T$ $[T_P]$ & $C_T$ $[C_P]$ &
$P_T$ $[P_P]$ & $S_T$ $[S_P]$ & $P_{EW,T}$ $[P_{EW,P}]$ &
$P_{EW,T}^C$ $[P_{EW,P}^C]$
\\ \hline
$B^+ \rightarrow \pi^+ a_2^0$ & $-{1 \over \sqrt{2}}$ & 1 & [1] &
$1, [-1]$ & 0 & [1] & $2 \over 3$, $\left[ 1 \over 3 \right] $
\\ $B^+ \rightarrow \pi^+ f_2$ & $1 \over \sqrt{2}$ & $c$ & $[c]$ & $c, [c]$ & $[2c + \sqrt{2} s]$ &
$\left[ {1 \over 3} (c- \sqrt{2} s) \right]$ & $2c \over 3$,
$\left[ -{c \over 3} \right]$
\\ $B^+ \rightarrow \pi^+ f_2^{\prime}$ & $1 \over \sqrt{2}$ & $s$ & $[s]$ & $s, [s]$ & $[2s - \sqrt{2} c]$ &
$\left[ {1 \over 3} (s+ \sqrt{2} c) \right]$ & $2s \over 3$,
$\left[ -{s \over 3} \right]$
\\ $B^+ \rightarrow \pi^0 a_2^+$ & $-{1 \over \sqrt{2}}$ & [1] & 1 & $-1$, [1] & 0 & 1 & ${1 \over 3}$,
$\left[ 2 \over 3 \right]$
\\ $B^+ \rightarrow \eta a_2^+$ & $-{1 \over \sqrt{3}}$ & [1] & 1 & 1, [1] & 1 & $2 \over 3$ &
$-{1 \over 3}$, $\left[ 2 \over 3 \right]$
\\ $B^+ \rightarrow \eta^{\prime} a_2^+$ & $1 \over \sqrt{6}$ & [1] & 1 & 1, [1] & 4 & $-{1 \over 3}$ &
$-{1 \over 3}$, $\left[ 2 \over 3 \right]$
\\ $B^+ \rightarrow K^+ \bar K_2^{* 0}$ & 1 & 0 & 0 & [1] & 0 & 0 & $\left[ -{1 \over 3} \right]$
\\ $B^+ \rightarrow \bar K^0 K_2^{* +}$ & 1 & 0 & 0 & 1 & 0 & 0 & $-{1 \over 3}$
\\  $B^0 \rightarrow \pi^+ a_2^-$ & $-1$ & 1 & 0 & 1 & 0 & 0 & $2 \over 3$
\\ $B^0 \rightarrow \pi^- a_2^+$ & $-1$ & [1] & 0 & [1] & 0 & 0 & $\left[ 2 \over 3 \right]$
\\ $B^0 \rightarrow \pi^0 a_2^0$ & ${1 \over 2}$ & 0 & $-1$, $[-1]$ & 1, [1] & 0 & $-1$, $[-1]$ &
$-{1 \over 3}$, $\left[ -{1 \over 3} \right]$
\\ $B^0 \rightarrow \pi^0 f_2$ & $-{1 \over 2}$ & 0 & $c$, $[-c]$ & $-c$, $[-c]$ & $[-(2c + \sqrt{2} s)]$ &
$c$, $[-{1 \over 3}(c -\sqrt{2} s)]$ & $c \over 3$, $\left[ c
\over 3 \right]$
\\ $B^0 \rightarrow \pi^0 f_2^{\prime}$ & $-{1 \over 2}$ & 0 & $s$, $[-s]$ & $-s$, $[-s]$ & $[-(2s - \sqrt{2} c)]$ &
$s$, $[-{1 \over 3}(s +\sqrt{2} c)]$ & $s \over 3$, $\left[ s
\over 3 \right]$
\\ $B^0 \rightarrow \eta a_2^0$ & $1 \over \sqrt{6}$ & 0 & $-1$, $[1]$ & $-1$, $[-1]$ & $-1$ & $-{2 \over 3}$, $[1]$ &
${1 \over 3}$, $\left[ {1 \over 3} \right]$
\\ $B^0 \rightarrow \eta f_2$ & $-{1 \over \sqrt{6}}$ & 0 & $c$, $[c]$ & $c$, $[c]$ & $c$, $[2c +\sqrt{2} s]$ &
$2c \over 3$, $\left[ {1 \over 3} (c -\sqrt{2} s) \right]$ & $-{c
\over 3}$, $\left[ -{c \over 3} \right]$
\\ $B^0 \rightarrow \eta f_2^{\prime}$ & $-{1 \over \sqrt{6}}$ & 0 & $s$, $[s]$ & $s$, $[s]$ & $s$, $[2s -\sqrt{2} c]$ &
$2s \over 3$, $\left[ {1 \over 3} (s +\sqrt{2} c) \right]$ & $-{s
\over 3}$, $\left[ -{s \over 3} \right]$
\\ $B^0 \rightarrow \eta^{\prime} a_2^0$ & $-{1 \over 2\sqrt{3}}$ & 0 & $-1$, $[1]$ & $-1$, $[-1]$ & $-4$ &
${1 \over 3}$, $[1]$ & ${1 \over 3}$, $[{1 \over 3}]$
\\ $B^0 \rightarrow \eta^{\prime} f_2$ & $1 \over 2\sqrt{3}$ & 0 & $c$, $[c]$ & $c$, $[c]$ & $4c$, $[2c +\sqrt{2} s]$ &
$-{c \over 3}$, $\left[ {1 \over 3} (c -\sqrt{2} s) \right]$ &
$-{c \over 3}$, $[-{c \over 3}]$
\\ $B^0 \rightarrow \eta^{\prime} f_2^{\prime}$ & $1 \over 2\sqrt{3}$ & 0 & $s$, $[s]$ & $s$, $[s]$ &
$4s$, $[2s -\sqrt{2} c]$ & $-{s \over 3}$, $\left[ {1 \over 3} (s
+\sqrt{2} c) \right]$ & $-{s \over 3}$, $[-{s \over 3}]$
\\ $B^0 \rightarrow K^0 \bar K_2^{* 0}$ & 1 & 0 & 0 & [1] & 0 & 0 & $\left[ -{1 \over 3} \right]$
\\ $B^0 \rightarrow \bar K^0 K_2^{* 0}$ & 1 & 0 & 0 & 1 & 0 & 0 & $-{1 \over 3}$
\end{tabular}
\end{table}

\begin{table}
\caption{Coefficients of SU(3) amplitudes in $B \rightarrow PT$
($|\Delta S| = 1$).}
\begin{tabular}{c||c|c|c|c|c|c|c}
$B \rightarrow PT$ & factor & $T_T^{\prime}$ $[T_P^{\prime}]$ &
$C_T^{\prime}$ $[C_P^{\prime}]$ & $P_T^{\prime}$ $[P_P^{\prime}]$
& $S_T^{\prime}$ $[S_P^{\prime}]$ & $P_{EW,T}^{\prime}$
$[P_{EW,P}^{\prime}]$ & $P_{EW,T}^{C \prime}$ $[P_{EW,P}^{C
\prime}]$
\\ \hline
$B^+ \rightarrow K^+ a_2^0$ & $-{1 \over \sqrt{2}}$ & 1 & [1] & 1
& 0 & [1] & $2 \over 3$
\\ $B^+ \rightarrow K^+ f_2$ & $1 \over \sqrt{2}$ & $c$ & $[c]$ & $c$, $[\sqrt{2} s]$ & $[2 c + \sqrt{2} s]$ &
$\left[ {1 \over 3} (c -\sqrt{2} s) \right]$ & $2c \over 3$,
$\left[ -{\sqrt{2} s \over 3} \right]$
\\ $B^+ \rightarrow K^+ f_2^{\prime}$ & $1 \over \sqrt{2}$ & $s$ & $[s]$ & $s$, $[-\sqrt{2} c]$ & $[2 s -\sqrt{2} c]$ &
$\left[ {1 \over 3} (s +\sqrt{2} c) \right]$ & $2s \over 3$,
$\left[ {\sqrt{2} c \over 3} \right]$
\\ $B^+ \rightarrow K^0 a_2^+$ & 1 & 0 & 0 & 1 & 0 & 0 & $-{1 \over 3}$
\\ $B^+ \rightarrow \pi^+ K_2^{* 0}$ & 1 & 0 & 0 & [1] & 0 & 0 & $\left[ -{1 \over 3} \right]$
\\ $B^+ \rightarrow \pi^0 K_2^{* +}$ & $-{1 \over \sqrt{2}}$ & [1] & 1 & [1] & 0 & 1 & $\left[ {2 \over 3} \right]$
\\ $B^+ \rightarrow \eta K_2^{* +}$ & $-{1 \over \sqrt{3}}$ & [1] & 1 & $-1$, [1] & 1 & ${2 \over 3}$ &
${1 \over 3}$, $\left[ {2 \over 3} \right]$
\\ $B^+ \rightarrow \eta^{\prime} K_2^{* +}$ & ${1 \over \sqrt{6}}$ & [1] & 1 & 2, [1] & 4 & $-{1 \over 3}$ &
$-{2 \over 3}$, $\left[ {2 \over 3} \right]$
\\  $B^0 \rightarrow K^+ a_2^-$ & $-1$ & 1 & 0 & 1 & 0 & 0 & ${2 \over 3}$
\\ $B^0 \rightarrow K^0 a_2^0$ & $-{1 \over \sqrt{2}}$ & 0 & [1] & $-1$ & 0 & [1] & ${1 \over 3}$
\\ $B^0 \rightarrow K^0 f_2$ & $1 \over \sqrt{2}$ & 0 & $[c]$ & $c$, $[\sqrt{2} s]$ & $[2 c + \sqrt{2} s]$ &
$\left[ {1 \over 3} (c -\sqrt{2} s) \right]$ & $-{c \over 3}$,
$\left[ -{\sqrt{2} s \over 3} \right]$
\\ $B^0 \rightarrow K^0 f_2^{\prime}$ & $1 \over \sqrt{2}$ & 0 & $[s]$ & $s$, $[-\sqrt{2} c]$ & $[2 s -\sqrt{2} c]$ &
$\left[ {1 \over 3} (s +\sqrt{2} c) \right]$ & $-{s \over 3}$,
$\left[ {\sqrt{2} c \over 3} \right]$
\\ $B^0 \rightarrow \pi^- K_2^{* +}$ & $-1$ & [1] & 0 & [1] & 0 & 0 & $\left[ {2 \over 3} \right]$
\\ $B^0 \rightarrow \pi^0 K_2^{* 0}$ & $-{1 \over \sqrt{2}}$ & 0 & 1 & $[-1]$ & 0 & 1 & $\left[ {1 \over 3} \right]$
\\ $B^0 \rightarrow \eta K_2^{* 0}$ & $-{1 \over \sqrt{3}}$ & 0 & 1 & $-1$, [1] & 1 & ${2 \over 3}$ &
${1 \over 3}$, $\left[ -{1 \over 3} \right]$
\\ $B^0 \rightarrow \eta^{\prime} K_2^{* 0}$ & ${1 \over \sqrt{6}}$ & 0 & 1 & 2, [1] & 4 & $-{1 \over 3}$ &
$-{2 \over 3}$, $\left[ -{1 \over 3} \right]$
\end{tabular}
\end{table}

\begin{table}
\caption{The branching ratios for $B \rightarrow PT$ decay modes
with $\Delta S =0$. The second and the third columns correspond to
the cases of sets of the parameters: \{$\xi =0.1$, $m_s = 85$ MeV,
$\gamma =110^0$\} and \{$\xi =0.1$, $m_s = 100$ MeV, $\gamma
=65^0$\}, respectively. Similarly, the fourth and the fifth
columns corresponds to the cases: \{$\xi =0.3$, $m_s = 85$ MeV,
$\gamma =110^0$\} and \{$\xi =0.3$, $m_s = 100$ MeV, $\gamma
=65^0$\}, respectively.  The sixth and the seventh columns
correspond to the cases: \{$\xi =0.5$, $m_s = 85$ MeV, $\gamma
=110^0$\} and \{$\xi =0.5$, $m_s = 100$ MeV, $\gamma =65^0$\},
respectively.}
\smallskip
\begin{tabular}{c|cccccc}

Decay mode                                      & ${\cal
B}(10^{-8})$& ${\cal B}(10^{-8})$&${\cal B}(10^{-8})$ &${\cal
B}(10^{-8})$ &  ${\cal B}(10^{-8})$ & ${\cal B}(10^{-8})$

\\ \hline
   $B^+ \rightarrow \pi^+ a_2^0$               &45.41  &44.82       &40.32   & 39.82     &35.54  &35.11
\\ $B^+ \rightarrow \pi^+ f_2$                 &49.31  &48.67       &43.79  &43.24       &38.59  &38.13
\\ $B^+ \rightarrow \pi^+ f_2^{\prime}$        &0.46  &0.46         &0.41   & 0.40        &0.36  &0.36
\\ $B^+ \rightarrow \pi^0 a_2^+$               &1.78  &1.52           &0.029   &0.048     &2.05   &2.38
\\ $B^+ \rightarrow \eta a_2^+$                &5.81  &6.02           &5.20    &3.94     &7.09  &4.48
\\ $B^+ \rightarrow \eta^{\prime} a_2^+$       &27.19 &22.97         &23.02    &17.93    &20.33  &14.45

\\ $B^+ \rightarrow \bar K^0 K_2^{* +}$        &0.025  &0.013         &0.032   & 0.019    &0.041  &0.026
\\
   $B^0 \rightarrow \pi^+ a_2^-$                &85.91  &84.80       &76.29   &75.34       &67.23 &66.44
\\ $B^0 \rightarrow \pi^0 a_2^0$                &0.84  &0.72          &0.014    &0.023    &0.97  &1.12
\\ $B^0 \rightarrow \pi^0 f_2$                  &0.92  &0.78          &0.015    &0.025     &1.05  &1.22
\\ $B^0 \rightarrow \pi^0 f_2^{\prime}$         &0.009  &0.007         &0.0001   & 0.0001   &0.010  &0.011
\\ $B^0 \rightarrow \eta a_2^0$                 &2.75 &2.85          &2.46   & 1.86          &3.36  &2.12
\\ $B^0 \rightarrow \eta f_2$                   &2.99  &3.09          &2.67   &2.02          &3.65  &2.30
\\ $B^0 \rightarrow \eta f_2^{\prime}$          &0.03  &0.03           &0.025    &0.019      &0.024  &0.021
\\ $B^0 \rightarrow \eta^{\prime} a_2^0$        &12.86  &10.87        &10.89    &8.48        &9.62 &6.83
\\ $B^0 \rightarrow \eta^{\prime} f_2$          &14.00  &10.87         &11.85   &9.23          &10.47  &7.44
\\ $B^0 \rightarrow \eta^{\prime} f_2^{\prime}$ &0.13  &0.11          &0.11    &0.085          &0.096  &0.068

\\$B^0 \rightarrow \bar K^0 K_2^{* 0}$          &0.023  &0.012             &0.030   &0.017    &0.038  &0.024

\end{tabular}
\end{table}


\begin{table}
\caption{The CP asymmetries for $B \rightarrow PT$ decay modes
with $\Delta S =0$.  The definitions for the columns are the same
as those in Table III. }
\smallskip
\begin{tabular}{c|cccccc}

Decay mode                  &${\cal A_{CP}}$ &${\cal A_{CP}}$
&${\cal A_{CP}}$& ${\cal A_{CP}}$ & ${\cal A_{CP}}$& ${\cal
A_{CP}}$

\\ \hline
   $B^+ \rightarrow \pi^+ a_2^0$                &0.016  &0.016      & 0.015  & 0.015       &0.015  &0.015
\\ $B^+ \rightarrow \pi^+ f_2$                  &0.016  &0.016      &0.015   &0.015        &0.015  &0.015
\\ $B^+ \rightarrow \pi^+ f_2^{\prime}$         &0.016  &0.016       & 0.015& 0.015         &0.015 &0.015
\\ $B^+ \rightarrow \pi^0 a_2^+$                &0.14  &0.15         & $-0.89$   &$-0.52$   &$-0.13$  &$-0.10$
\\ $B^+ \rightarrow \eta a_2^+$                 &0.59 &0.55         &$-0.068$  &$-0.087$    &$-0.46$ & $-0.71$
\\ $B^+ \rightarrow \eta^{\prime} a_2^+$        &0.17  &0.20        &$-0.021$    &$-0.026$   &$-0.22$ & $-0.29$

\\ $B^+ \rightarrow \bar K^0 K_2^{* +}$         &$0 $  & $0$         &0   &0                &0  &0
\\
   $B^0 \rightarrow \pi^+ a_2^-$                &0.016  &0.015        & 0.015  &0.015       &0.015  &0.015
\\ $B^0 \rightarrow \pi^0 a_2^0$                &0.14  &0.15          &$-0.89$  &$-0.52$   &$-0.13$  &$-0.10$
\\ $B^0 \rightarrow \pi^0 f_2$                  &0.14 &0.15          &$-0.89$  &$-0.52$    &$-0.13$  &$-0.10$
\\ $B^0 \rightarrow \pi^0 f_2^{\prime}$         &0.14  &0.14         & $-0.89$  &$-0.52$     &$-0.13$  &$-0.10$
\\ $B^0 \rightarrow \eta a_2^0$                 &0.59 &0.55          &$-0.068$  &$-0.087$    &$-0.46$  &$-0.71$
\\ $B^0 \rightarrow \eta f_2$                   &0.59 &0.59          &$-0.068$   &$-0.087$     &$-0.46$  &$-0.71$
\\ $B^0 \rightarrow \eta f_2^{\prime}$          &0.59  &0.55         &$-0.068$   &$-0.087$  &$-0.46$  &$-0.71$
\\ $B^0 \rightarrow \eta^{\prime} a_2^0$        &0.17 &0.20         &$-0.021$   &$-0.026$        &0.22 & $-0.29$
\\ $B^0 \rightarrow \eta^{\prime} f_2$          &0.17  &0.20         & $-0.021$  &$-0.026$       &$-0.22$  &$-0.29$
\\ $B^0 \rightarrow \eta^{\prime} f_2^{\prime}$ &0.17 &0.20          &$-0.021$   &$-0.026$         &$-0.22$ & $-0.29$

\\$B^0 \rightarrow \bar K^0 K_2^{* 0}$          &$0$ &0     &0    &0                             &$0$  &0

\end{tabular}
\end{table}

\newpage
\begin{table}
\caption{The branching ratios  for $B \rightarrow PT$ decay modes
with $|\Delta S| =1$.  The definitions for the columns are the
same as those in Table III.}
\smallskip
\begin{tabular}{c|cccccc}

Decay mode                                     & ${\cal
B}(10^{-8})$& ${\cal B}(10^{-8})$&${\cal B}(10^{-8})$ &${\cal
B}(10^{-8})$& ${\cal B}(10^{-8})$ & ${\cal B}(10^{-8})$

\\ \hline
   $B^+ \rightarrow K^+ a_2^0$                 &4.31   &5.77         & 3.81   & 5.08      &3.34  &4.43
\\ $B^+ \rightarrow K^+ f_2$                   &4.69 &6.27           &4.14    & 5.52       &3.63 &4.82
\\ $B^+ \rightarrow K^+ f_2^{\prime}$          &0.044  &0.058        & 3.84   & 0.051      &0.037 &0.045
\\ $B^+ \rightarrow K^0 a_2^+$                 &5.08 &1.22           & 6.22 & 1.97          &7.47  &2.91

\\ $B^+ \rightarrow \pi^0 K_2^{* +}$           &1.13  &1.55          & 1.09  & 1.05      &1.19  &0.75
\\ $B^+ \rightarrow \eta K_2^{* +}$            &0.10  &0.23          & 0.035  &0.22       &0.077 &0.31
\\ $B^+ \rightarrow \eta^{\prime} K_2^{* +}$   &43.09  &26.58        & 44.96  &29.98       &46.91 &33.64
\\
   $B^0 \rightarrow K^+ a_2^-$                 &8.16  &10.92 & 7.21   & 9.61 &6.32  &8.39
\\ $B^0 \rightarrow K^0 a_2^0$                 &2.40&0.58 &2.94  & 0.93 &3.53  &1.38
\\ $B^0 \rightarrow K^0 f_2$                   &2.61 &0.63  & 3.20  & 1.01 &3.84  &1.50
\\ $B^0 \rightarrow K^0 f_2^{\prime}$          &0.024  &0.006 & 0.030 & 0.009 &0.036 &0.014

\\ $B^0 \rightarrow \pi^0 K_2^{* 0}$           &1.05 &1.45  & 1.02  & 0.98  &1.11  &0.70
\\ $B^0 \rightarrow \eta K_2^{* 0}$            &0.095 &0.21 & 0.033 & 0.21 &0.072 &0.29
\\ $B^0 \rightarrow \eta^{\prime} K_2^{* 0}$   &40.14 &24.76 &41.88 &27.93 &43.70 &31.34

\end{tabular}
\end{table}

\begin{table}
\caption{The CP asymmetries for $B \rightarrow PT$ decay modes
with $|\Delta S| =1$.  The definitions for the columns are the
same as those in Table III.}
\smallskip
\begin{tabular}{c|cccccc}

Decay mode                                      &${\cal A_{CP}}$
&${\cal A_{CP}}$ &${\cal A_{CP}}$& ${\cal A_{CP}}$ & ${\cal
A_{CP}}$& ${\cal A_{CP}}$

\\ \hline
   $B^+ \rightarrow K^+ a_2^0$                  &$-0.11$   &0.022         & $-0.11$ & 0.022           &$-0.11$ &0.022
\\ $B^+ \rightarrow K^+ f_2$                    &$-0.12$ &0.022             & $-0.11$  & 0.022         &$-0.11$&0.022
\\ $B^+ \rightarrow K^+ f_2^{\prime}$           &$-0.12$  &0.022        & $-0.11 $& 0.022              & $-0.11$ &0.022
\\ $B^+ \rightarrow K^0 a_2^+$                  &0 &0                    & 0   & 0                     &0  &0

\\ $B^+ \rightarrow \pi^0 K_2^{* +}$            &0.006  &0.004          & $-0.001$  & $-0.001$      &$-0.007$&$-0.010$
\\ $B^+ \rightarrow \eta K_2^{* +}$            &0.65  &0.39           & $-0.21$ & $-0.043$           &$-0.92$ &$-0.31$
\\ $B^+ \rightarrow \eta^{\prime} K_2^{* +}$    &0.005  &0.006        & $-0.001$  & $-0.001$         & $-0.005$&$-0.005$
\\
   $B^0 \rightarrow K^+ a_2^-$                 &$-0.12$  &0.022      &$-0.11$   & 0.022         &$-0.11$  &0.022
\\ $B^0 \rightarrow K^0 a_2^0$                 &0  &0                 & 0  & 0                    &0  &0
\\ $B^0 \rightarrow K^0 f_2$                   &0  &0                & 0  & 0                    &0  &0
\\ $B^0 \rightarrow K^0 f_2^{\prime}$           &0  &0               & 0 & 0                  &0  &0

\\ $B^0 \rightarrow \pi^0 K_2^{* 0}$            &0.006  &0.004       & $-0.001$  & $-0.001$    &$-0.007$ &$-0.010$
\\ $B^0 \rightarrow \eta K_2^{* 0}$            &0.65  &0.39          & $-0.21$ & $-0.043$        &$-0.92$  &$-0.31$
\\ $B^0 \rightarrow \eta^{\prime} K_2^{* 0}$    &0.005  &0.006       &$-0.001$   & $-0.001$         &$-0.005$ &$-0.005$

\end{tabular}
\end{table}
\end{document}